\documentclass[aps,prl,twocolumn,superscriptaddress,floatfix]{revtex4}

\usepackage{bm}
\usepackage{amssymb}
\usepackage{graphicx}
\usepackage{amsmath}
\usepackage{dsfont}
\usepackage{dcolumn}
\usepackage[pdftex]{epsfig}
\usepackage{color}
\usepackage[german,english]{babel}
\usepackage{psfrag}
\usepackage{hyperref}
\usepackage[caption=false]{subfig}
\usepackage{longtable}
\usepackage{psfrag}
\usepackage{color}
\usepackage{float}

\definecolor{myRed}{rgb}{0.8, 0.2, 0.2}

\newcommand{\mv}[1]{\mathbf{#1}}
\renewcommand{\Im}{\operatorname{Im}}

\begin{document}

\title{Valley Plasmonics in the  Dichalcogenides}

\author{R. E. Groenewald}
\thanks{R. E. Groenewald and M. R{\"o}sner contributed equally to this work.}
\affiliation{Department of Physics and Astronomy, University of Southern California, Los Angeles, CA 90089-0484, USA}

\author{M. R{\"o}sner}
\thanks{R. E. Groenewald and M. R{\"o}sner contributed equally to this work.}
\affiliation{Institut f{\"u}r Theoretische Physik, Universit{\"a}t Bremen, Otto-Hahn-Allee 1, 28359 Bremen, Germany}
\affiliation{Bremen Center for Computational Materials Science, Universit{\"a}t Bremen, Am Fallturm 1a, 28359 Bremen, Germany}

\author{G. Sch{\"o}nhoff}
\affiliation{Institut f{\"u}r Theoretische Physik, Universit{\"a}t Bremen, Otto-Hahn-Allee 1, 28359 Bremen, Germany}
\affiliation{Bremen Center for Computational Materials Science, Universit{\"a}t Bremen, Am Fallturm 1a, 28359 Bremen, Germany}

\author{S. Haas}
\affiliation{Department of Physics and Astronomy, University of Southern California, Los Angeles, CA 90089-0484, USA}

\author{T. O. Wehling}
\affiliation{Institut f{\"u}r Theoretische Physik, Universit{\"a}t Bremen, Otto-Hahn-Allee 1, 28359 Bremen, Germany}
\affiliation{Bremen Center for Computational Materials Science, Universit{\"a}t Bremen, Am Fallturm 1a, 28359 Bremen, Germany}

\date{\today}

\begin{abstract}
	The rich phenomenology of plasmonic excitations in the dichalcogenides is analyzed as a function of doping. The many-body polarization, the dielectric response function and electron energy loss spectra are calculated using an ab initio based model involving material-realistic Coulomb interactions, band structure and spin-orbit coupling. Focusing on the representative case of $\rm MoS_2$, a plethora of plasmon bands are observed, originating from scattering processes within and between the conduction or valence band valleys. We discuss the resulting square-root and linear collective modes, arising from long-range versus short-range screening of the Coulomb potential. We show that the multi-orbital nature of the bands and spin-orbit coupling strongly affects \emph{inter-valley} scattering processes by gapping certain two-particle modes at large momentum transfer. 
\end{abstract}


\maketitle

{\it Introduction:}
	Collective excitations are of great interest in low-dimensional materials which are characterized by reduced dielectric screening of Coulomb interactions. As a prominent example, plasmon modes in layered systems might form the basis to build optical devices, wave guides or so called plasmonic circuits \cite{koppens_graphene_2011, bao_graphene_2012, grigorenko_graphene_2012, garcia_de_abajo_graphene_2014}. In two dimensions (2d) the plasmonic dispersion exhibits a characteristic low-energy acoustic mode $\omega(q) \propto \sqrt q$ originating from low-momentum electron scattering \cite{ritchie_plasma_1957, bhukal_dispersion_2015}, which has been observed experimentally \cite{liu_plasmon_2008, ju_graphene_2011} and studied extensively from a theoretical point of view \cite{wunsch_dynamical_2006, hwang_dielectric_2007, gangadharaiah_charge_2008, grigorenko_graphene_2012, stauber_plasmonics_2014} in graphene. Furthermore, it has been predicted that additional linear plasmons with $\omega(q) \propto q$  arise due to high-momentum scattering processes between degenerated valleys such as $K$ and $K'$ in graphene \cite{tudorovskiy_intervalley_2010}. Coupling of the electrons with such intrinsic gapless bosonic modes may lead to instabilities, such as charge density wave and superconducting phases \cite{rietschel_role_1983, bill_electronic_2003, akashi_development_2013, linscheid_ab_2015}, similar to the effect of phonons. 

	An analogous but even richer phenomenology can be expected in the structurally related monolayer transition metal dichalcogenides (TMDCs) MX$_2$, where M stands for a transition metal and X for a chalcogen atom. These materials host rich plasmonic physics including an interplay of plasmons with charge density waves \cite{van_wezel_effect_2011, konig_plasmon_2012, konig_doping_2013} and first plasmon based applications have already been proposed \cite{kalantar-zadeh_two-dimensional_2015, kalantar-zadeh_biosensors_2015, maurya_performance_2015}.

\begin{figure}[ht]
		\includegraphics[width=0.9\linewidth]{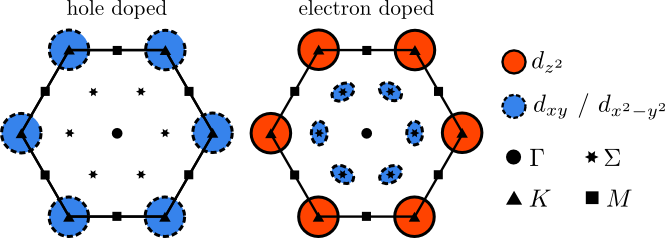}
		\caption{(Color online) Sketch of the Fermi surfaces in hole (left) and electron doped (right) monolayer MoS$_2$. The different orbital characters are indicated by red ($d_{z^2}$) and blue ($d_{xy}$ and $d_{x^2-y^2}$) filled surfaces. Points of high symmetry are indicated by different markers. 
		}
		\label{fig:FS}
	\end{figure}

	Here we focus on the representative example of doped $\rm MoS_2$ whose low-energy band structure can be described by three effective tight-binding bands. These originate from the Mo $d$ orbitals, giving rise to prominent valleys at wave vectors $K$ and $\Sigma$ in the lowest conduction band as well as at $K$ in the highest valence band, leading to Fermi surfaces as depicted in Fig. \ref{fig:FS}. Furthermore, there is substantial spin-orbit coupling (SOC) in these materials \cite{zhu_giant_2011}, with a primary effect on the low-energy physics by introducing a splitting of the $\Sigma$ valleys in the lowest conduction band and of the $K$ valleys in the highest valence band. Although all of these Fermi surface characteristics can be experimentally sampled by means of field effect electron or hole doping \cite{chu_charge_2014}, the resulting impact to the plasmonic dispersions is not known.
	
	To close this gap, we present in this letter an extensive study of the plasmon dispersion at arbitrary momenta along paths throughout the whole Brillouin zone for different doping levels. Specifically we are interested in \emph{inter-valley} plasmons which have not been studied in TMDCs so far. In order to highlight the multi orbital character of the Fermi surface (see Fig. \ref{fig:FS}) and the presence of spin-orbit coupling we consider hole and electron doped cases. In the hole doped example we show how spin-orbit coupling affects the \emph{inter-valley} plasmons while the electron doped case is used to study the influence of the multi pocket structure of the Fermi surface. Thereby we gain a comprehensive and realistic picture of the most important contributions to the \emph{low energy} plasmon modes in monolayer TMDCs. \\

	\begin{figure*}[ht]
		\begin{tabular}{ccc}
			\includegraphics[width=0.3\linewidth]{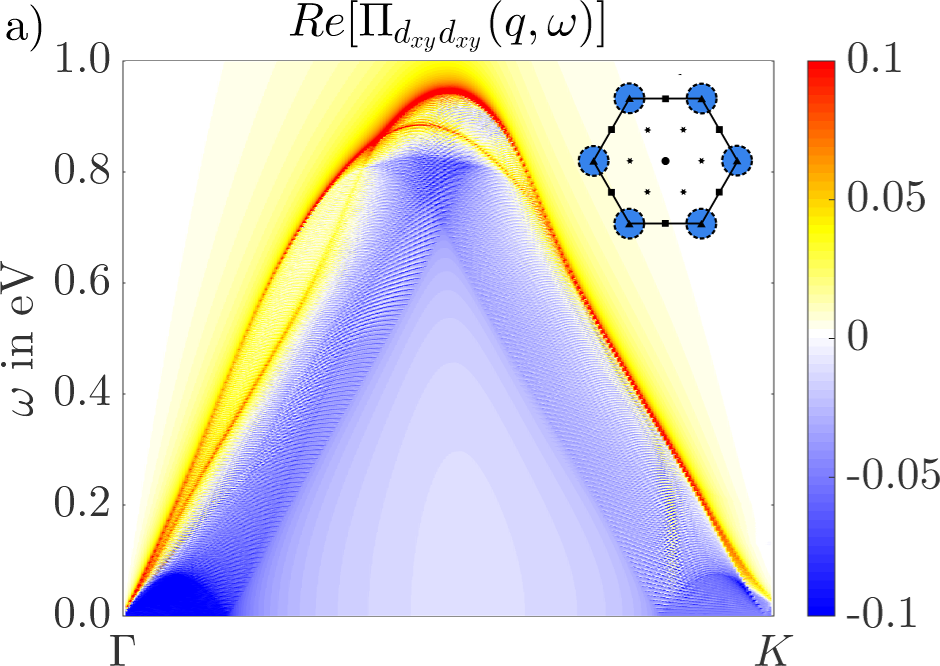}		
			&
			\includegraphics[width=0.3\linewidth]{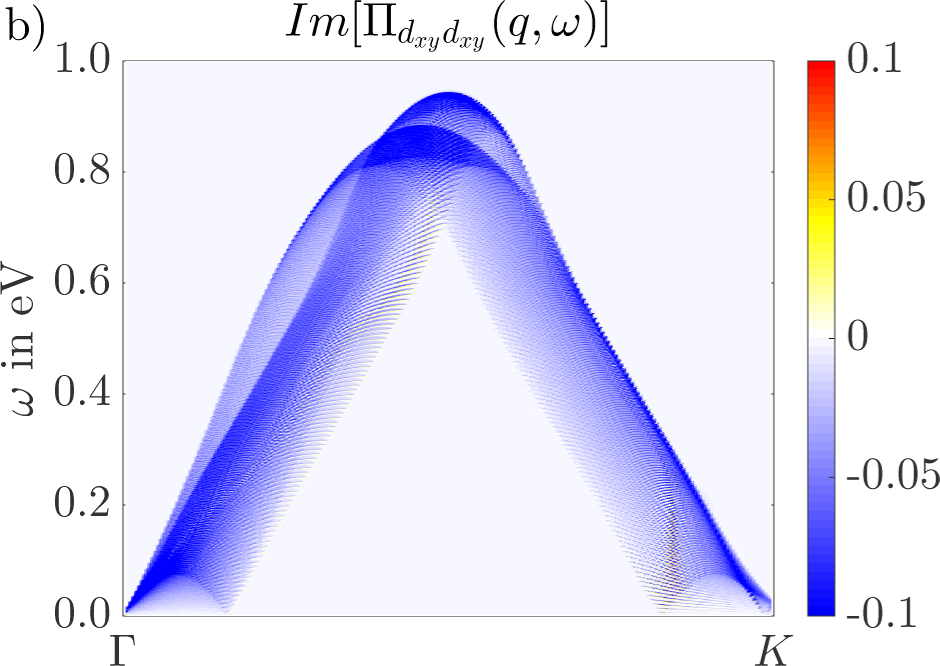}
			&
			\includegraphics[width=0.3\linewidth]{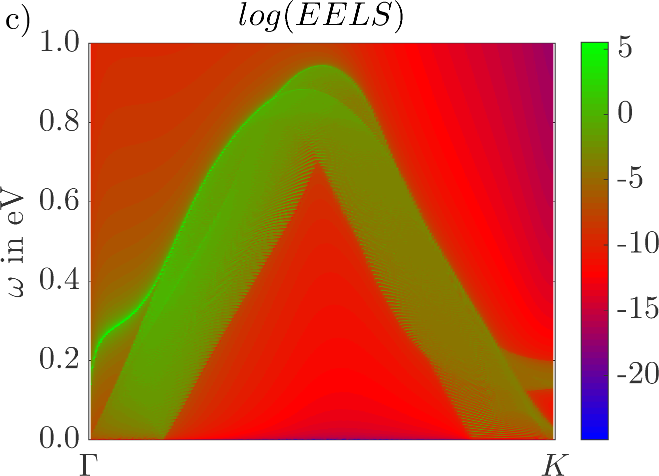}
		\\
			\includegraphics[width=0.3\linewidth]{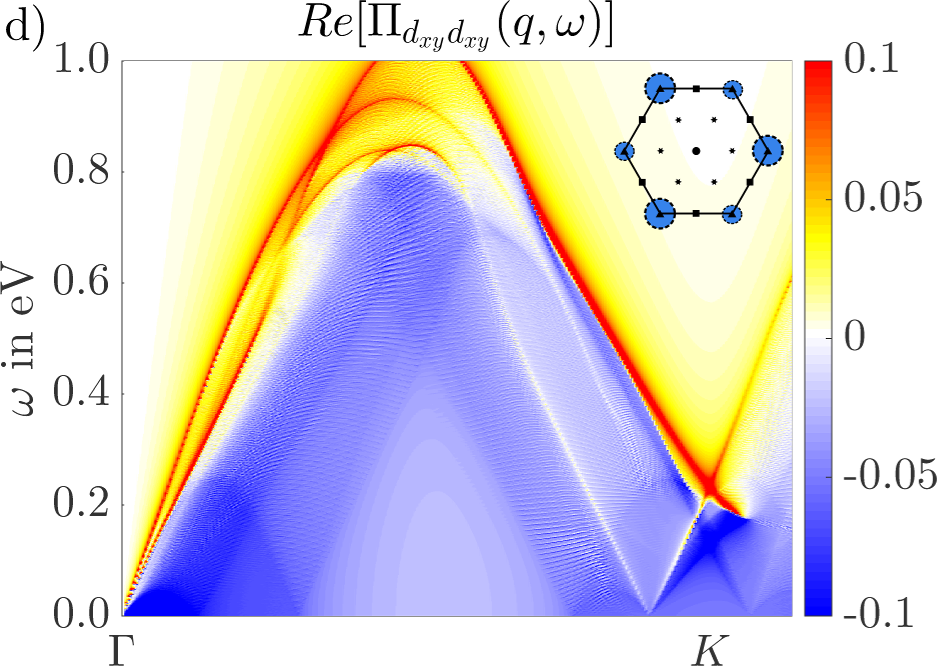} 
			&
			\includegraphics[width=0.3\linewidth]{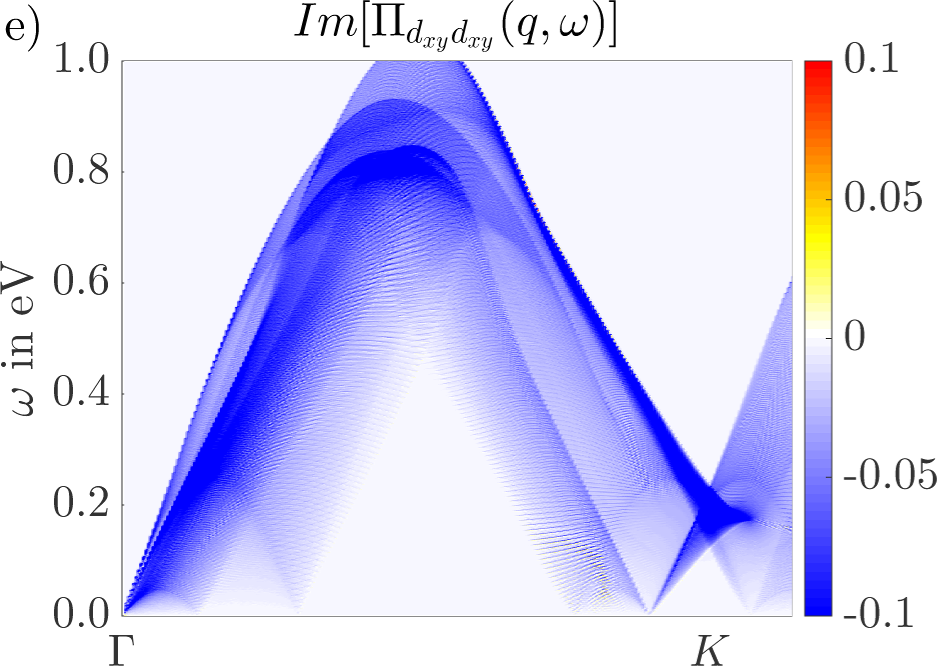} 
			&
			\includegraphics[width=0.3\linewidth]{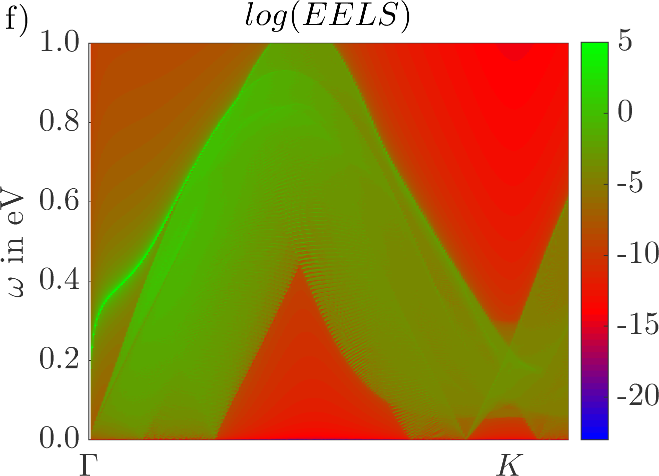}
		\end{tabular}
		\caption{(Color online) Real and imaginary parts of the polarization functions ($d_{xy}/d_{xy}$ channel) and EELS spectra for hole doped MoS$_2$ without (top row) and with (bottom row) spin orbit coupling. The insets in (a) and (d) illustrate the Fermi surface pockets around $K$ and $K'$. \label{fig:holeDope}}
	\end{figure*}

{\it Method:}
	The collective plasmon modes are described by the polarization and dielectric functions, which we evaluate in several steps, starting with a $G_0W_0$ calculation to determine the electronic band structure for the \emph{undoped} system. We then obtain an effective 3-band model by projecting to a Wannier basis spanned by the Mo $d_{x^2-y^2}$, $d_{xy}$ and $d_{z^2}$ orbitals, which has been found to accurately describe the highest valence band and the two lowest conduction bands with tight-binding hopping matrix elements $t_{\alpha\beta}$, where $\alpha$ and $\beta$ are the orbital indices. The same projection is used to obtain the static part of the Coulomb interaction in the Wannier basis, which is screened by \emph{all} bands including those, which are not included in the minimal 3-band-model
	\footnote{
		In all ab initio calculations we used an interlayer separation of $35\,$\AA. The $G_0W_0$ calculations are performed with the Vienna ab initio simulation package (VASP) \cite{kresse_textitab_1993, kresse_efficiency_1996}, while the Coulomb matrix elements are obtained from the SPEX code \cite{friedrich_efficient_2010} with FLAPW input from the FLEUR code \cite{_juelich_2014} as described in \cite{steinhoff_influence_2014}. For the involved Wannier projections we use the Wannier90 package \cite{mostofi_wannier90:_2008}. See supplemental material for further details.
	}.
	
	This procedure leads to an effective material-specific model with screened Coulomb $U_{\alpha\beta\gamma\delta}$ and hopping $t_{\alpha\beta}$ matrix elements in the orbital basis, describing the undoped system in its ground state. We find that this treatment is essential to derive material realistic plasmonic dispersions upon doping. In contrast to simplified $k \cdot p$ models \cite{scholz_plasmons_2013, kechedzhi_weakly_2014}, which utilize \emph{bare} Coulomb matrix elements at this stage, our interaction matrix elements are strongly reduced due to screening effects from the electronic bands which are neglected in the $k \cdot p$ models. As a result of the 2d layer geometry, these dielectric properties cannot be modeled by a simple dielectric \emph{constant} but have to be described as a $\mv{q}$-dependent dielectric \emph{function} \cite{steinhoff_influence_2014, liang_carrier_2015}. 
	 
	In order to obtain the dynamic response in the \emph{doped} system, we determine the dynamic susceptibility within the 3-orbital basis by evaluating the polarization in the random phase approximation (RPA), which is given for a single spin channel $\sigma$ by
	\begin{equation}
		\Pi^\sigma_{\alpha\beta}({\bf q},\omega) = 
			\sum_{\lambda_1\lambda_2\mv{k}}
				\frac
					{M_{\alpha\beta}^{\lambda_1\lambda_2} \left[f^\sigma_{\lambda_2}(\mv{k}+\mv{q}) - f^\sigma_{\lambda_1}(\mv{k})\right]}
					{\omega + i\delta + E^\sigma_{\lambda_2}(\mv{k}+\mv{q}) - E^\sigma_{\lambda_1}(\mv{k}) }, \label{eqn:PolRPA}
	\end{equation}
	where $\mv{q}$ and $\mv{k}$ are wave vectors from the first Brillouin zone, $\lambda_i$ band indices, $f^\sigma_{\lambda_i}(\mv{k})$ Fermi functions for the energies $E^\sigma_{\lambda_i}(\mv{k})$ and $i\delta$ a small broadening parameter. The overlap matrix elements are given by $M_{\alpha\beta}^{\lambda_1\lambda_2} = \bar{c}^{\lambda_1}_\alpha(\mv{k}) c^{\lambda_1}_\beta(\mv{k}) \bar{c}^{\lambda_2}_\beta(\mv{k} + \mv{q}) c^{\lambda_2}_\alpha(\mv{k} + \mv{q})$, where $c^{\lambda_i}_\alpha(\mv{k})$ is the expansion coefficient of the eigenfunction corresponding to $E^\sigma_{\lambda_i}(\mv{k})$ in the orbital basis. Here, we already reduced the polarization tensor of 4\textsuperscript{th} order to a matrix to describe density-density correlations only. Hence, we neglect orbital exchange (Fock-like) matrix elements as well as elements with three or even four different orbital contributions. A detailed analysis of the full background screened Coulomb tensor $U_{\alpha\beta\gamma\delta}$ shows, that these elements are in general one order of magnitude smaller or even vanish due to symmetries, which convinces us to stay with density-density like elements. 
	
	Using the full density-density polarization $\Pi(\mv{q}, \omega) = \Pi^\uparrow(\mv{q}, \omega) + \Pi^\downarrow(\mv{q}, \omega)$ the dielectric function is obtained via the following matrix equation
	\begin{equation}
		\varepsilon(\mv{q}, \omega)= \mathds{1} - U(\mv{q}) \Pi(\mv{q},\omega ) , \label{eqn:EpsRPA}
	\end{equation}
	where the background screened Coulomb interaction enters via $U(\mv{q})$. By including an effective spin-orbit coupling \cite{liu_three-band_2013} the spin degeneracy is removed but time reversal symmetry is preserved. Then, the spin resolved band structure still obeys $E_\lambda^\uparrow(\mv{k}) = E_\lambda^\downarrow(\mv{-k})$ and the total polarization including the spin summation can be written as $\Pi(\mv{q}, \omega) = \Pi^\uparrow(\mv{q}, \omega) + \Pi^\uparrow(-\mv{q}, \omega)$.
	
	The dielectric function describes the \emph{screened} Coulomb matrix $V(\mv{q}, \omega) = \varepsilon^{-1}(\mv{q}, \omega) U(\mv{q})$ and implicitly defines the plasmonic dispersions by $\varepsilon_m(\mv{q}, \omega) = 0$, where $\varepsilon_m$ is the \emph{macroscopic} part of the dielectric function \cite{rosner_wannier_2015}.	The most promising experimental method to map these plasmon modes is electron energy loss spectroscopy (EELS), measuring the imaginary part of the inverse dielectric function
	\begin{equation}
		\rm EELS({\bf q},\omega)=-\Im \left(\frac{1}{\varepsilon_m({\bf q},\omega)}\right),
	\end{equation}
	which is sensitive to both collective and single-particle excitations (visible as maxima in the EELS spectra) \cite{roth_electron_2014}.
	
	The combination of our material realistic description of the undoped system and the very accurate band structure for the RPA evaluation yields indeed quite accurate plasmon dispersions compared to full ab initio results, as we show for NbS$_2$ in the supplement. \\

{\it Hole doped $\rm MoS_2$:}
	We fix the chemical potential such that there are holes in the valence band in the $K$ and $K'$ valleys only. The resulting Fermi surfaces consists of circle-like areas around the $K$ points (see Fig. \ref{fig:FS}), which have mainly $d_{xy} / d_{x^2-y^2}$ character  and depend on spin-orbit coupling. Hence, we expect low energy plasmon modes for $\mv{q} \approx \mv{\Gamma}$ (intra-valley) and $\mv{q} \approx \mv{K}$ (inter-valley), which are possibly influenced by SOC.	 

	In Fig. \ref{fig:holeDope} (a) we show an intensity plot of the real part of the polarization function for scattering within $d_{xy}$ orbitals along the complete path $\mv{\Gamma} \rightarrow \mv{K}$ without SOC \footnote{
		Here we apply $\Gamma$ centered Monkhorst-Pack $720 \times 720$ $k$-grids and use a broadening of $i\delta = 0.0005i$. The doping concentration is adjusted by rigid shifts of the Fermi energy, which change the Fermi functions accordingly. All calculations are carried out for $T = 0\,$K. Furthermore, we restricted the $\lambda_1$ and $\lambda_2$ summations to the partially occupied band only in order to avoid double counting problems within the definition of the total polarization function and not to overload the resulting plots.
	}. Next to some band-like structures (red) we clearly see the particle-hole continuum (blue). In comparison to the corresponding EELS data in Fig. \ref{fig:holeDope} (c), we see that for higher momentum transfers (away from $\bf \Gamma$) the EELS maxima closely follow the band-like characteristics of the polarization function. For small momenta around $\bf \Gamma$ we find a clearly separated band in the EELS spectra, which can not be seen in the real part of the polarization. This separated band arises from the well known $\sqrt{q}$-dispersive \emph{intra-valley} plasmon mode in 2d \cite{scholz_plasmons_2013}, while we find a linear-dispersive mode around $\bf K$ stemming from an \emph{inter-valley} plasmon \cite{tudorovskiy_intervalley_2010}. These activation laws are consistent with the generalized expression for the plasmon dispersion relation defined by the dielectric function via \cite{bill_electronic_2003},
	\begin{equation}
		\omega(q) = 
			\hbar v_F q 
			\sqrt{ 1 + 
							\frac{ [N_0 U(q)]^2 }
							     { (1/4)+N_0 U(q) }
					 },
	\end{equation}
	where $v_F$ is the Fermi velocity, $N_0$ the density of states at the Fermi level and $U(q)$ the macroscopic background screened Coulomb interaction of the undoped system. In the long-wavelength limit ($q\rightarrow 0$),  the Coulomb potential remains unscreened, i.e. in leading order $U \propto 1/q$, resulting in a square-root renormalization of the otherwise linear dispersion. However, in the opposite short-range limit, i.e. at the zone boundary, the screened Coulomb potential approaches a constant, and therefore the resulting dispersion of the dielectric function is linear in $q$, same as the polarization function itself. Thus for momenta away from $\bf\Gamma$ it is sufficient to study the polarization function to understand how the resulting plasmon dispersion will behave.

	\begin{figure*}[ht]
		\includegraphics[width=0.3\linewidth]{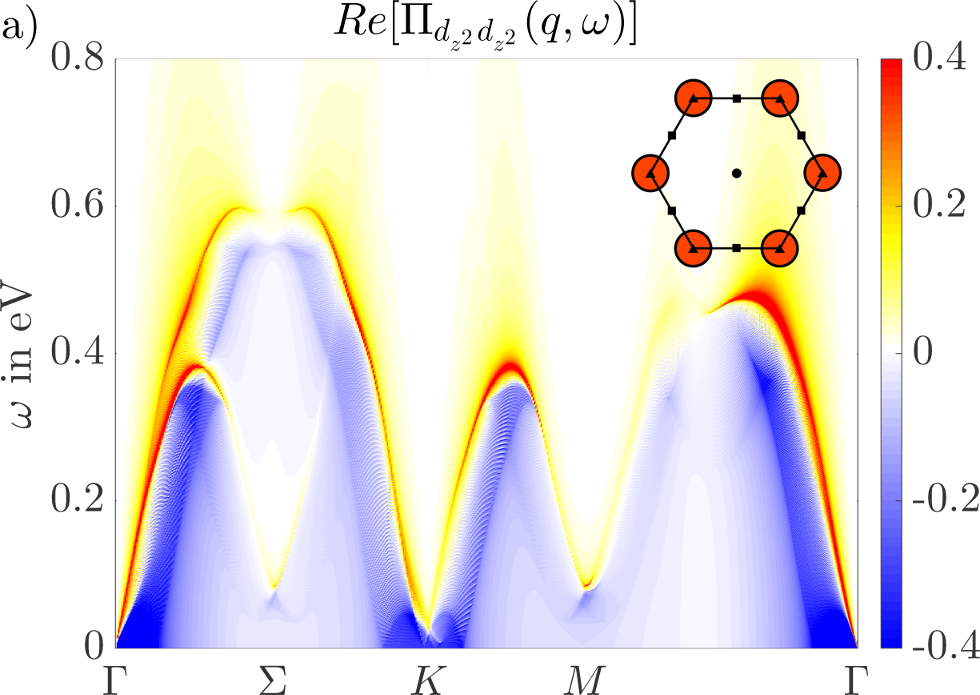}
		\includegraphics[width=0.3\linewidth]{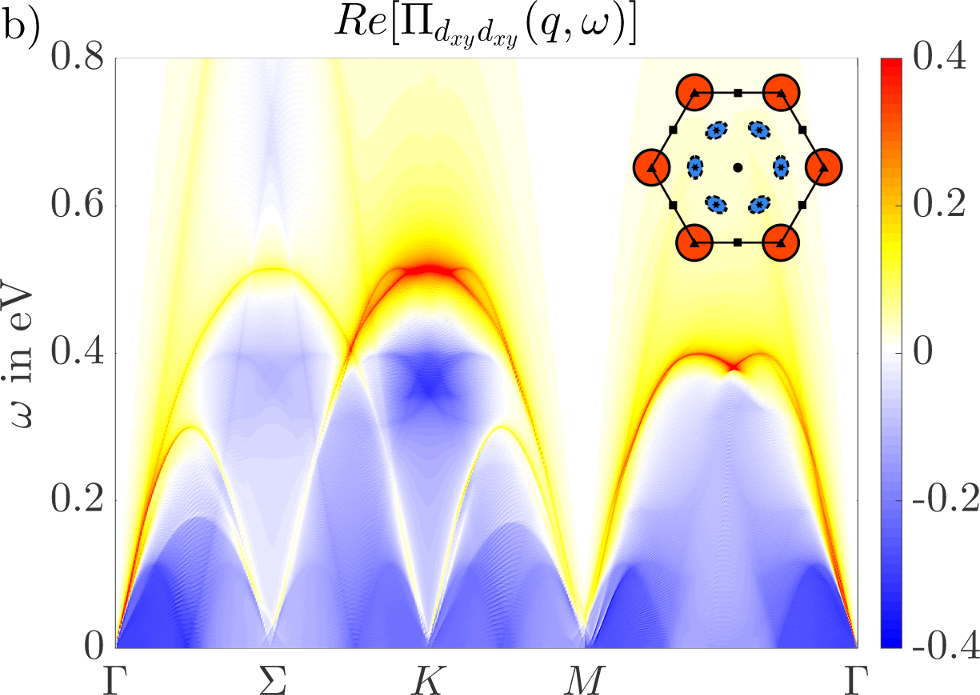}
		\includegraphics[width=0.3\linewidth]{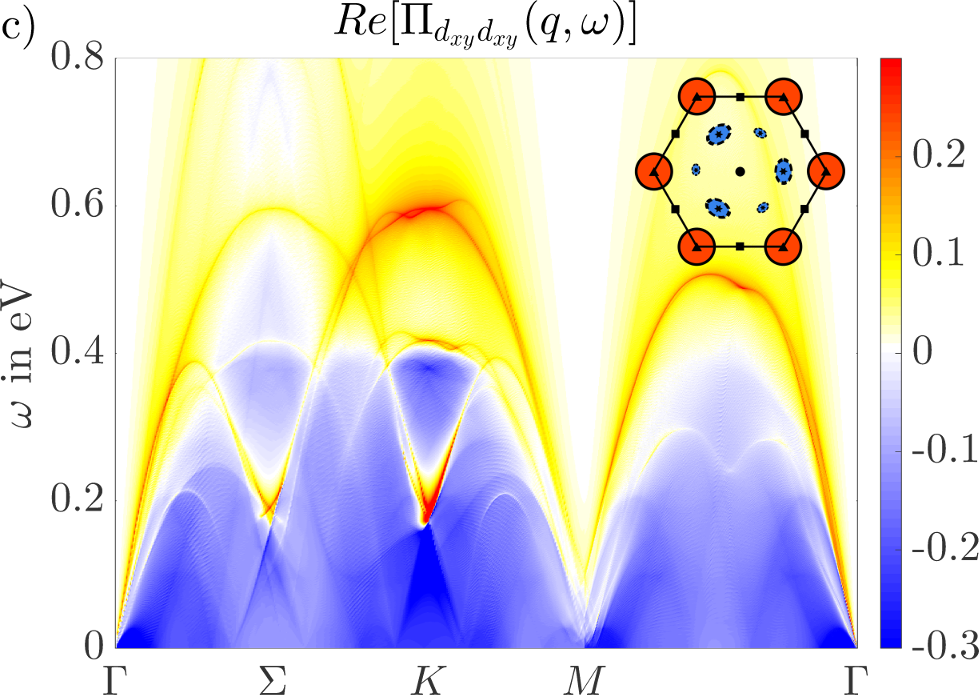}
		\caption{(Color online) (a) The polarization function for $d_{z^2}/d_{z^2}$ scattering at low electron doping concentration (only $K$ valleys are partially occupied) without SOC. (b) The polarization function for $d_{xy}/d_{xy}$ scattering at elevated electron doping concentration ($K$ and $\Sigma$ valleys are partially occupied) without SOC. In (c) the same situation as in (b) is shown, but with the effect of spin-orbit interaction. The insets show illustrations of the Fermi surfaces. \label{fig:electronDoped}}
	\end{figure*}
	
	Of special interest are damping effects, which are known to attenuate plasmon modes which merge with the particle-hole continuum. Here the square-root mode around $\bf \Gamma$ behaves in a distinctly different manner compared to the linear modes originating at $\bf K$. At sufficiently small momentum transfers $q < q_c$ the square-root modes are more separated from the nearby particle-hole continua [Fig. \ref{fig:holeDope} (c) and (f)], and therefore better protected from decomposition via hybridization and Landau damping [expressed as non-vanishing imaginary parts of the polarization as shown in Fig. \ref{fig:holeDope} (b) and (e)] compared to the linear modes originating at finite momenta. In contrast, the linear plasmon modes are much closer to their neighboring continua [Fig. \ref{fig:holeDope} (c)], which leads to attenuation effects, reflected in reduced oscillator strength and broadening of the peaks. There is a significant difference in the oscillator strengths of these modes, which can be several orders of magnitude apart as can be seen in Fig. \ref{fig:holeDope} (c) and (f). Hence, in order to clearly detect these linear plasmon modes in experiments, it may prove practical to use a logarithmic scale to shield the dominant square-root mode around $\mv{q} = \mv{\Gamma}$, as shown in Fig. \ref{fig:holeDope} (c) and (f).
	
	When we account for spin-orbit coupling the relative depth of the $K$ and $K'$ pockets shifts. In this case momentum transfer of $\mv{q} = \mv{K}$ no longer connects points on the Fermi surface belonging to different hole pockets, which results in two clearly visible characteristics in the polarization of Fig. \ref{fig:holeDope} (d): (1) At $\mv{q} = \mv{K}$ the scattering process is possible only for a finite energy difference, which opens a finite energy gap of $\approx 250\,$meV. (2) The Fermi surfaces at $K$ and $K'$ are now of different sizes but can still be connected with slightly smaller and larger $\mv{q}$, resulting in gap-less linear modes originating slightly shifted from $K$ as seen in Fig. \ref{fig:holeDope} (d). 
	
	We conclude that the plasmonic features in hole doped MoS$_2$ are qualitatively similar to graphene as long as SOC is not taken into account and the $K$ valley is occupied solely. Upon inclusion of SOC the linear plasmon mode around $K$ is \emph{shifted} leading to a gapped excitation spectra at this point.

{\it Electron doped $\rm MoS_2$:}
	The lowest conduction band is characterized by two prominent minima around $K$ and $\Sigma$. Without SOC these minima are separated by only $90\,$meV. Hence, in contrast to the hole doped case, small variations in the electron doping can change the Fermi surface drastically. In order to study these changes, we will neglect the SOC for the beginning and choose two doping levels, resulting in Fermi surfaces comparable to the hole doped case (i.e. $K$ valley occupation only) and a surface with additional pockets at $\Sigma$, labeled by low- and high-doping respectively (see Supplement). Since the $K$ valley is described by $d_{z^2}$ orbitals and the $\Sigma$ valley predominately by $d_{xy}$ and $d_{x^2-y^2}$ states, we focus on corresponding diagonal orbital channels in $\Pi_{\alpha\beta}$ in the following. Off-diagonal elements between $d_{z^2}$ and $d_{xy}/d_{x^2-y^2}$ orbitals are negligible here (off-diagonal terms between $d_{xy}/d_{x^2-y^2}$ states are similar). The corresponding polarization functions are shown along the path $\Gamma-\Sigma-K-M-\Gamma$ through the whole Brillouin zone in Fig. \ref{fig:electronDoped}. 
	
	Analogous to the hole doped case, we observe around $\bf q = \Gamma$ the expected resonances arising from \emph{intra-valley} scattering. This is naturally present in both high and low electron doping cases. By inspection of Fig. \ref{fig:FS} we can understand the structure of the polarization for larger $\bf q$. 
	
	The momenta $\bf q = \Sigma, K$ and $\bf M$ connect different $\Sigma$ valleys. Therefore, in the high electron doping case (with $\Sigma$ partially occupied) we expect additional \emph{inter-valley} plasmon branches close to these momenta.
	
	At $\bf q \approx K$ we observe plasmon bands in both, high and low doping cases since this momentum transfer allows inter-valley scattering between $K$ and $\Sigma$ pockets. As we are calculating orbital resolved polarization functions, the observed low energy excitation in Fig. \ref{fig:electronDoped} (a) is due to $K \leftrightarrow K'$ and thus $d_{z^2}$ scattering, whereas in Fig. \ref{fig:electronDoped} (b) it is due to $\Sigma \leftrightarrow \Sigma$ and correspondingly $d_{xy}/d_{x^2-y^2}$ scattering ($K \leftrightarrow K'$ scattering is obviously still present, but can only be seen in the $d_{z^2}$ polarization as shown in the Supplement).
	
	Finally, momentum transfers $\bf q = M$ and $\bf \Sigma$ can connect different $\Sigma$ valleys, and therefore we find a gap-less linear \emph{inter-valley} plasmon mode originating at this point only in the high doping case. In the low doping case, we observe a gapped ($\approx 0.1eV$) excitation at $\bf q = M$, originating from a $K \leftrightarrow \Sigma$ excitations. 

	While the SOC has a negligible effect on the $d_{z^2}$ valley at $K$ it splits the $d_{xy} / d_{x^2-y^2}$ valleys at $\Sigma$ resulting in minima at comparable energies. The corresponding Fermi surface for a single spin component is indicated in the inset of Fig. \ref{fig:electronDoped} (c). The six $\Sigma$ points decompose into two distinct sets, $\Sigma$ and $\Sigma'$. Fermi pockets within each of these subsets are mutually connected by $2\pi/3$ rotations and remain equivalent after inclusion of SOC, while the degeneracy of $\Sigma$ and $\Sigma'$ is lifted by SOC. As a consequence, the phase space for $\Sigma \leftrightarrow \Sigma'$ is lost and the gap-less excitations at $\bf q \approx \Sigma$ and $\bf q \approx K$ must vanish, but $\Sigma \leftrightarrow \Sigma$ scattering processes are still possible. Consequently, we see in the corresponding polarization for the $d_{xy}$ channel with SOC in Fig. \ref{fig:electronDoped} (c) gap-less modes only at $\bf \Gamma$ and $\bf M$. Since the Fermi surface around $K$ is not changed drastically upon SOC, the corresponding polarization for the $d_{z^2}$ channel is very similar to the one obtained without SOC (see Supplement).

{\it Conclusions:} 
	We found that the low energy dynamical screening in MoS$_2$ is controlled by both \emph{inter-} and \emph{intra-valley} scattering processes. These give rise to plasmons with a square root dispersion at small $\bf q$ and linear dispersion for higher momentum transfers which connect separate valleys on the Fermi surface. In general, inter-valley plasmon modes are observable, although their oscillator strengths are strongly reduced in comparison to zone center modes. Due to the multi-orbital character of the wave functions and spin-orbit coupling, which leads to spin-valley coupling in monolayer TMDCs, not all \emph{inter-valley} scattering processes are allowed. As a consequence of spin-valley coupling some inter-valley plasmon modes are shifted and gapped out, while the $2\pi/3$ rotation symmetry protects certain low energy modes at $\bf M$. We speculate this selective gapping out of collective modes could have consequences for the realization of many-body instabilities towards superconducting or charge density wave phases in monolayer TMDCs.

{\it Acknowledgments:} 
	We are grateful for useful discussions with A.V. Balatsky, A. Bill, F. Guinea as well as B. Normand. S.H. would like to the thank the Humboldt Foundation for support. This work was supported by the European Graphene Flagship and by the Department of Energy under Grant No. DE-FG02-05ER46240. The numerical computations were carried out on the University of Southern California high performance supercomputer cluster and the Norddeutscher Verbund zur F\"{o}rderung des Hoch- und H\"{o}chstleistungsrechnens (HLRN) cluster. 

\bibliography{references}

\begin{thebibliography}{37}%
\makeatletter
\providecommand \@ifxundefined [1]{%
 \@ifx{#1\undefined}
}%
\providecommand \@ifnum [1]{%
 \ifnum #1\expandafter \@firstoftwo
 \else \expandafter \@secondoftwo
 \fi
}%
\providecommand \@ifx [1]{%
 \ifx #1\expandafter \@firstoftwo
 \else \expandafter \@secondoftwo
 \fi
}%
\providecommand \natexlab [1]{#1}%
\providecommand \enquote  [1]{``#1''}%
\providecommand \bibnamefont  [1]{#1}%
\providecommand \bibfnamefont [1]{#1}%
\providecommand \citenamefont [1]{#1}%
\providecommand \href@noop [0]{\@secondoftwo}%
\providecommand \href [0]{\begingroup \@sanitize@url \@href}%
\providecommand \@href[1]{\@@startlink{#1}\@@href}%
\providecommand \@@href[1]{\endgroup#1\@@endlink}%
\providecommand \@sanitize@url [0]{\catcode `\\12\catcode `\$12\catcode
  `\&12\catcode `\#12\catcode `\^12\catcode `\_12\catcode `\%12\relax}%
\providecommand \@@startlink[1]{}%
\providecommand \@@endlink[0]{}%
\providecommand \url  [0]{\begingroup\@sanitize@url \@url }%
\providecommand \@url [1]{\endgroup\@href {#1}{\urlprefix }}%
\providecommand \urlprefix  [0]{URL }%
\providecommand \Eprint [0]{\href }%
\providecommand \doibase [0]{http://dx.doi.org/}%
\providecommand \selectlanguage [0]{\@gobble}%
\providecommand \bibinfo  [0]{\@secondoftwo}%
\providecommand \bibfield  [0]{\@secondoftwo}%
\providecommand \translation [1]{[#1]}%
\providecommand \BibitemOpen [0]{}%
\providecommand \bibitemStop [0]{}%
\providecommand \bibitemNoStop [0]{.\EOS\space}%
\providecommand \EOS [0]{\spacefactor3000\relax}%
\providecommand \BibitemShut  [1]{\csname bibitem#1\endcsname}%
\let\auto@bib@innerbib\@empty
\bibitem [{\citenamefont {Koppens}\ \emph {et~al.}(2011)\citenamefont
  {Koppens}, \citenamefont {Chang},\ and\ \citenamefont {Garc{\'i}a~de
  Abajo}}]{koppens_graphene_2011}%
  \BibitemOpen
  \bibfield  {author} {\bibinfo {author} {\bibfnamefont {F.~H.~L.}\
  \bibnamefont {Koppens}}, \bibinfo {author} {\bibfnamefont {D.~E.}\
  \bibnamefont {Chang}}, \ and\ \bibinfo {author} {\bibfnamefont {F.~J.}\
  \bibnamefont {Garc{\'i}a~de Abajo}},\ }\href {\doibase 10.1021/nl201771h}
  {\bibfield  {journal} {\bibinfo  {journal} {Nano Letters}\ }\textbf {\bibinfo
  {volume} {11}},\ \bibinfo {pages} {3370} (\bibinfo {year}
  {2011})}\BibitemShut {NoStop}%
\bibitem [{\citenamefont {Bao}\ and\ \citenamefont
  {Loh}(2012)}]{bao_graphene_2012}%
  \BibitemOpen
  \bibfield  {author} {\bibinfo {author} {\bibfnamefont {Q.}~\bibnamefont
  {Bao}}\ and\ \bibinfo {author} {\bibfnamefont {K.~P.}\ \bibnamefont {Loh}},\
  }\href {\doibase 10.1021/nn300989g} {\bibfield  {journal} {\bibinfo
  {journal} {ACS Nano}\ }\textbf {\bibinfo {volume} {6}},\ \bibinfo {pages}
  {3677} (\bibinfo {year} {2012})}\BibitemShut {NoStop}%
\bibitem [{\citenamefont {Grigorenko}\ \emph {et~al.}(2012)\citenamefont
  {Grigorenko}, \citenamefont {Polini},\ and\ \citenamefont
  {Novoselov}}]{grigorenko_graphene_2012}%
  \BibitemOpen
  \bibfield  {author} {\bibinfo {author} {\bibfnamefont {A.~N.}\ \bibnamefont
  {Grigorenko}}, \bibinfo {author} {\bibfnamefont {M.}~\bibnamefont {Polini}},
  \ and\ \bibinfo {author} {\bibfnamefont {K.~S.}\ \bibnamefont {Novoselov}},\
  }\href {\doibase 10.1038/nphoton.2012.262} {\bibfield  {journal} {\bibinfo
  {journal} {Nature Photonics}\ }\textbf {\bibinfo {volume} {6}},\ \bibinfo
  {pages} {749} (\bibinfo {year} {2012})}\BibitemShut {NoStop}%
\bibitem [{\citenamefont {Garc{\'i}a~de
  Abajo}(2014)}]{garcia_de_abajo_graphene_2014}%
  \BibitemOpen
  \bibfield  {author} {\bibinfo {author} {\bibfnamefont {F.~J.}\ \bibnamefont
  {Garc{\'i}a~de Abajo}},\ }\href {\doibase 10.1021/ph400147y} {\bibfield
  {journal} {\bibinfo  {journal} {ACS Photonics}\ }\textbf {\bibinfo {volume}
  {1}},\ \bibinfo {pages} {135} (\bibinfo {year} {2014})}\BibitemShut {NoStop}%
\bibitem [{\citenamefont {Ritchie}(1957)}]{ritchie_plasma_1957}%
  \BibitemOpen
  \bibfield  {author} {\bibinfo {author} {\bibfnamefont {R.~H.}\ \bibnamefont
  {Ritchie}},\ }\href {\doibase 10.1103/PhysRev.106.874} {\bibfield  {journal}
  {\bibinfo  {journal} {Physical Review}\ }\textbf {\bibinfo {volume} {106}},\
  \bibinfo {pages} {874} (\bibinfo {year} {1957})}\BibitemShut {NoStop}%
\bibitem [{\citenamefont {Bhukal}\ \emph {et~al.}(2015)\citenamefont {Bhukal},
  \citenamefont {{Priya}},\ and\ \citenamefont
  {Moudgil}}]{bhukal_dispersion_2015}%
  \BibitemOpen
  \bibfield  {author} {\bibinfo {author} {\bibfnamefont {N.}~\bibnamefont
  {Bhukal}}, \bibinfo {author} {\bibnamefont {{Priya}}}, \ and\ \bibinfo
  {author} {\bibfnamefont {R.~K.}\ \bibnamefont {Moudgil}},\ }\href {\doibase
  10.1016/j.physe.2015.01.007} {\bibfield  {journal} {\bibinfo  {journal}
  {Physica E: Low-dimensional Systems and Nanostructures}\ }\textbf {\bibinfo
  {volume} {69}},\ \bibinfo {pages} {13} (\bibinfo {year} {2015})}\BibitemShut
  {NoStop}%
\bibitem [{\citenamefont {Liu}\ \emph {et~al.}(2008)\citenamefont {Liu},
  \citenamefont {Willis}, \citenamefont {Emtsev},\ and\ \citenamefont
  {Seyller}}]{liu_plasmon_2008}%
  \BibitemOpen
  \bibfield  {author} {\bibinfo {author} {\bibfnamefont {Y.}~\bibnamefont
  {Liu}}, \bibinfo {author} {\bibfnamefont {R.~F.}\ \bibnamefont {Willis}},
  \bibinfo {author} {\bibfnamefont {K.~V.}\ \bibnamefont {Emtsev}}, \ and\
  \bibinfo {author} {\bibfnamefont {T.}~\bibnamefont {Seyller}},\ }\href
  {\doibase 10.1103/PhysRevB.78.201403} {\bibfield  {journal} {\bibinfo
  {journal} {Physical Review B}\ }\textbf {\bibinfo {volume} {78}},\ \bibinfo
  {pages} {201403} (\bibinfo {year} {2008})}\BibitemShut {NoStop}%
\bibitem [{\citenamefont {Ju}\ \emph {et~al.}(2011)\citenamefont {Ju},
  \citenamefont {Geng}, \citenamefont {Horng}, \citenamefont {Girit},
  \citenamefont {Martin}, \citenamefont {Hao}, \citenamefont {Bechtel},
  \citenamefont {Liang}, \citenamefont {Zettl}, \citenamefont {Shen},\ and\
  \citenamefont {Wang}}]{ju_graphene_2011}%
  \BibitemOpen
  \bibfield  {author} {\bibinfo {author} {\bibfnamefont {L.}~\bibnamefont
  {Ju}}, \bibinfo {author} {\bibfnamefont {B.}~\bibnamefont {Geng}}, \bibinfo
  {author} {\bibfnamefont {J.}~\bibnamefont {Horng}}, \bibinfo {author}
  {\bibfnamefont {C.}~\bibnamefont {Girit}}, \bibinfo {author} {\bibfnamefont
  {M.}~\bibnamefont {Martin}}, \bibinfo {author} {\bibfnamefont
  {Z.}~\bibnamefont {Hao}}, \bibinfo {author} {\bibfnamefont {H.~A.}\
  \bibnamefont {Bechtel}}, \bibinfo {author} {\bibfnamefont {X.}~\bibnamefont
  {Liang}}, \bibinfo {author} {\bibfnamefont {A.}~\bibnamefont {Zettl}},
  \bibinfo {author} {\bibfnamefont {Y.~R.}\ \bibnamefont {Shen}}, \ and\
  \bibinfo {author} {\bibfnamefont {F.}~\bibnamefont {Wang}},\ }\href {\doibase
  10.1038/nnano.2011.146} {\bibfield  {journal} {\bibinfo  {journal} {Nature
  Nanotechnology}\ }\textbf {\bibinfo {volume} {6}},\ \bibinfo {pages} {630}
  (\bibinfo {year} {2011})}\BibitemShut {NoStop}%
\bibitem [{\citenamefont {Wunsch}\ \emph {et~al.}(2006)\citenamefont {Wunsch},
  \citenamefont {Stauber}, \citenamefont {Sols},\ and\ \citenamefont
  {Guinea}}]{wunsch_dynamical_2006}%
  \BibitemOpen
  \bibfield  {author} {\bibinfo {author} {\bibfnamefont {B.}~\bibnamefont
  {Wunsch}}, \bibinfo {author} {\bibfnamefont {T.}~\bibnamefont {Stauber}},
  \bibinfo {author} {\bibfnamefont {F.}~\bibnamefont {Sols}}, \ and\ \bibinfo
  {author} {\bibfnamefont {F.}~\bibnamefont {Guinea}},\ }\href {\doibase
  10.1088/1367-2630/8/12/318} {\bibfield  {journal} {\bibinfo  {journal} {New
  Journal of Physics}\ }\textbf {\bibinfo {volume} {8}},\ \bibinfo {pages}
  {318} (\bibinfo {year} {2006})}\BibitemShut {NoStop}%
\bibitem [{\citenamefont {Hwang}\ and\ \citenamefont
  {Das~Sarma}(2007)}]{hwang_dielectric_2007}%
  \BibitemOpen
  \bibfield  {author} {\bibinfo {author} {\bibfnamefont {E.~H.}\ \bibnamefont
  {Hwang}}\ and\ \bibinfo {author} {\bibfnamefont {S.}~\bibnamefont
  {Das~Sarma}},\ }\href {\doibase 10.1103/PhysRevB.75.205418} {\bibfield
  {journal} {\bibinfo  {journal} {Physical Review B}\ }\textbf {\bibinfo
  {volume} {75}},\ \bibinfo {pages} {205418} (\bibinfo {year}
  {2007})}\BibitemShut {NoStop}%
\bibitem [{\citenamefont {Gangadharaiah}\ \emph {et~al.}(2008)\citenamefont
  {Gangadharaiah}, \citenamefont {Farid},\ and\ \citenamefont
  {Mishchenko}}]{gangadharaiah_charge_2008}%
  \BibitemOpen
  \bibfield  {author} {\bibinfo {author} {\bibfnamefont {S.}~\bibnamefont
  {Gangadharaiah}}, \bibinfo {author} {\bibfnamefont {A.~M.}\ \bibnamefont
  {Farid}}, \ and\ \bibinfo {author} {\bibfnamefont {E.~G.}\ \bibnamefont
  {Mishchenko}},\ }\href {\doibase 10.1103/PhysRevLett.100.166802} {\bibfield
  {journal} {\bibinfo  {journal} {Physical Review Letters}\ }\textbf {\bibinfo
  {volume} {100}},\ \bibinfo {pages} {166802} (\bibinfo {year}
  {2008})}\BibitemShut {NoStop}%
\bibitem [{\citenamefont {Stauber}(2014)}]{stauber_plasmonics_2014}%
  \BibitemOpen
  \bibfield  {author} {\bibinfo {author} {\bibfnamefont {T.}~\bibnamefont
  {Stauber}},\ }\href {\doibase 10.1088/0953-8984/26/12/123201} {\bibfield
  {journal} {\bibinfo  {journal} {Journal of Physics: Condensed Matter}\
  }\textbf {\bibinfo {volume} {26}},\ \bibinfo {pages} {123201} (\bibinfo
  {year} {2014})}\BibitemShut {NoStop}%
\bibitem [{\citenamefont {Tudorovskiy}\ and\ \citenamefont
  {Mikhailov}(2010)}]{tudorovskiy_intervalley_2010}%
  \BibitemOpen
  \bibfield  {author} {\bibinfo {author} {\bibfnamefont {T.}~\bibnamefont
  {Tudorovskiy}}\ and\ \bibinfo {author} {\bibfnamefont {S.~A.}\ \bibnamefont
  {Mikhailov}},\ }\href {\doibase 10.1103/PhysRevB.82.073411} {\bibfield
  {journal} {\bibinfo  {journal} {Physical Review B}\ }\textbf {\bibinfo
  {volume} {82}},\ \bibinfo {pages} {073411} (\bibinfo {year}
  {2010})}\BibitemShut {NoStop}%
\bibitem [{\citenamefont {Rietschel}\ and\ \citenamefont
  {Sham}(1983)}]{rietschel_role_1983}%
  \BibitemOpen
  \bibfield  {author} {\bibinfo {author} {\bibfnamefont {H.}~\bibnamefont
  {Rietschel}}\ and\ \bibinfo {author} {\bibfnamefont {L.~J.}\ \bibnamefont
  {Sham}},\ }\href {\doibase 10.1103/PhysRevB.28.5100} {\bibfield  {journal}
  {\bibinfo  {journal} {Physical Review B}\ }\textbf {\bibinfo {volume} {28}},\
  \bibinfo {pages} {5100} (\bibinfo {year} {1983})}\BibitemShut {NoStop}%
\bibitem [{\citenamefont {Bill}\ \emph {et~al.}(2003)\citenamefont {Bill},
  \citenamefont {Morawitz},\ and\ \citenamefont
  {Kresin}}]{bill_electronic_2003}%
  \BibitemOpen
  \bibfield  {author} {\bibinfo {author} {\bibfnamefont {A.}~\bibnamefont
  {Bill}}, \bibinfo {author} {\bibfnamefont {H.}~\bibnamefont {Morawitz}}, \
  and\ \bibinfo {author} {\bibfnamefont {V.~Z.}\ \bibnamefont {Kresin}},\
  }\href {\doibase 10.1103/PhysRevB.68.144519} {\bibfield  {journal} {\bibinfo
  {journal} {Physical Review B}\ }\textbf {\bibinfo {volume} {68}},\ \bibinfo
  {pages} {144519} (\bibinfo {year} {2003})}\BibitemShut {NoStop}%
\bibitem [{\citenamefont {Akashi}\ and\ \citenamefont
  {Arita}(2013)}]{akashi_development_2013}%
  \BibitemOpen
  \bibfield  {author} {\bibinfo {author} {\bibfnamefont {R.}~\bibnamefont
  {Akashi}}\ and\ \bibinfo {author} {\bibfnamefont {R.}~\bibnamefont {Arita}},\
  }\href {\doibase 10.1103/PhysRevLett.111.057006} {\bibfield  {journal}
  {\bibinfo  {journal} {Physical Review Letters}\ }\textbf {\bibinfo {volume}
  {111}},\ \bibinfo {pages} {057006} (\bibinfo {year} {2013})}\BibitemShut
  {NoStop}%
\bibitem [{\citenamefont {Linscheid}\ \emph {et~al.}(2015)\citenamefont
  {Linscheid}, \citenamefont {Sanna},\ and\ \citenamefont
  {Gross}}]{linscheid_ab_2015}%
  \BibitemOpen
  \bibfield  {author} {\bibinfo {author} {\bibfnamefont {A.}~\bibnamefont
  {Linscheid}}, \bibinfo {author} {\bibfnamefont {A.}~\bibnamefont {Sanna}}, \
  and\ \bibinfo {author} {\bibfnamefont {E.~K.~U.}\ \bibnamefont {Gross}},\
  }\href {http://arxiv.org/abs/1503.00977} {\bibfield  {journal} {\bibinfo
  {journal} {arXiv:1503.00977 [cond-mat]}\ } (\bibinfo {year} {2015})},\
  \bibinfo {note} {arXiv: 1503.00977}\BibitemShut {NoStop}%
\bibitem [{\citenamefont {van Wezel}\ \emph {et~al.}(2011)\citenamefont {van
  Wezel}, \citenamefont {Schuster}, \citenamefont {K{\"o}nig}, \citenamefont
  {Knupfer}, \citenamefont {van~den Brink}, \citenamefont {Berger},\ and\
  \citenamefont {B{\"u}chner}}]{van_wezel_effect_2011}%
  \BibitemOpen
  \bibfield  {author} {\bibinfo {author} {\bibfnamefont {J.}~\bibnamefont {van
  Wezel}}, \bibinfo {author} {\bibfnamefont {R.}~\bibnamefont {Schuster}},
  \bibinfo {author} {\bibfnamefont {A.}~\bibnamefont {K{\"o}nig}}, \bibinfo
  {author} {\bibfnamefont {M.}~\bibnamefont {Knupfer}}, \bibinfo {author}
  {\bibfnamefont {J.}~\bibnamefont {van~den Brink}}, \bibinfo {author}
  {\bibfnamefont {H.}~\bibnamefont {Berger}}, \ and\ \bibinfo {author}
  {\bibfnamefont {B.}~\bibnamefont {B{\"u}chner}},\ }\href {\doibase
  10.1103/PhysRevLett.107.176404} {\bibfield  {journal} {\bibinfo  {journal}
  {Physical Review Letters}\ }\textbf {\bibinfo {volume} {107}},\ \bibinfo
  {pages} {176404} (\bibinfo {year} {2011})}\BibitemShut {NoStop}%
\bibitem [{\citenamefont {K{\"o}nig}\ \emph {et~al.}(2012)\citenamefont
  {K{\"o}nig}, \citenamefont {Koepernik}, \citenamefont {Schuster},
  \citenamefont {Kraus}, \citenamefont {Knupfer}, \citenamefont {B{\"u}chner},\
  and\ \citenamefont {Berger}}]{konig_plasmon_2012}%
  \BibitemOpen
  \bibfield  {author} {\bibinfo {author} {\bibfnamefont {A.}~\bibnamefont
  {K{\"o}nig}}, \bibinfo {author} {\bibfnamefont {K.}~\bibnamefont
  {Koepernik}}, \bibinfo {author} {\bibfnamefont {R.}~\bibnamefont {Schuster}},
  \bibinfo {author} {\bibfnamefont {R.}~\bibnamefont {Kraus}}, \bibinfo
  {author} {\bibfnamefont {M.}~\bibnamefont {Knupfer}}, \bibinfo {author}
  {\bibfnamefont {B.}~\bibnamefont {B{\"u}chner}}, \ and\ \bibinfo {author}
  {\bibfnamefont {H.}~\bibnamefont {Berger}},\ }\href {\doibase
  10.1209/0295-5075/100/27002} {\bibfield  {journal} {\bibinfo  {journal} {EPL
  (Europhysics Letters)}\ }\textbf {\bibinfo {volume} {100}},\ \bibinfo {pages}
  {27002} (\bibinfo {year} {2012})}\BibitemShut {NoStop}%
\bibitem [{\citenamefont {K{\"o}nig}\ \emph {et~al.}(2013)\citenamefont
  {K{\"o}nig}, \citenamefont {Schuster}, \citenamefont {Knupfer}, \citenamefont
  {B{\"u}chner},\ and\ \citenamefont {Berger}}]{konig_doping_2013}%
  \BibitemOpen
  \bibfield  {author} {\bibinfo {author} {\bibfnamefont {A.}~\bibnamefont
  {K{\"o}nig}}, \bibinfo {author} {\bibfnamefont {R.}~\bibnamefont {Schuster}},
  \bibinfo {author} {\bibfnamefont {M.}~\bibnamefont {Knupfer}}, \bibinfo
  {author} {\bibfnamefont {B.}~\bibnamefont {B{\"u}chner}}, \ and\ \bibinfo
  {author} {\bibfnamefont {H.}~\bibnamefont {Berger}},\ }\href {\doibase
  10.1103/PhysRevB.87.195119} {\bibfield  {journal} {\bibinfo  {journal}
  {Physical Review B}\ }\textbf {\bibinfo {volume} {87}},\ \bibinfo {pages}
  {195119} (\bibinfo {year} {2013})}\BibitemShut {NoStop}%
\bibitem [{\citenamefont {Kalantar-zadeh}\ \emph {et~al.}(2015)\citenamefont
  {Kalantar-zadeh}, \citenamefont {Ou}, \citenamefont {Daeneke}, \citenamefont
  {Strano}, \citenamefont {Pumera},\ and\ \citenamefont
  {Gras}}]{kalantar-zadeh_two-dimensional_2015}%
  \BibitemOpen
  \bibfield  {author} {\bibinfo {author} {\bibfnamefont {K.}~\bibnamefont
  {Kalantar-zadeh}}, \bibinfo {author} {\bibfnamefont {J.~Z.}\ \bibnamefont
  {Ou}}, \bibinfo {author} {\bibfnamefont {T.}~\bibnamefont {Daeneke}},
  \bibinfo {author} {\bibfnamefont {M.~S.}\ \bibnamefont {Strano}}, \bibinfo
  {author} {\bibfnamefont {M.}~\bibnamefont {Pumera}}, \ and\ \bibinfo {author}
  {\bibfnamefont {S.~L.}\ \bibnamefont {Gras}},\ }\href {\doibase
  10.1002/adfm.201500891} {\bibfield  {journal} {\bibinfo  {journal} {Advanced
  Functional Materials}\ }\textbf {\bibinfo {volume} {25}},\ \bibinfo {pages}
  {5086} (\bibinfo {year} {2015})}\BibitemShut {NoStop}%
\bibitem [{\citenamefont {Kalantar-zadeh}\ and\ \citenamefont
  {Ou}(2015)}]{kalantar-zadeh_biosensors_2015}%
  \BibitemOpen
  \bibfield  {author} {\bibinfo {author} {\bibfnamefont {K.}~\bibnamefont
  {Kalantar-zadeh}}\ and\ \bibinfo {author} {\bibfnamefont {J.~Z.}\
  \bibnamefont {Ou}},\ }\href {\doibase 10.1021/acssensors.5b00142} {\bibfield
  {journal} {\bibinfo  {journal} {ACS Sensors}\ } (\bibinfo {year} {2015}),\
  10.1021/acssensors.5b00142}\BibitemShut {NoStop}%
\bibitem [{\citenamefont {Maurya}\ \emph {et~al.}(2015)\citenamefont {Maurya},
  \citenamefont {Prajapati}, \citenamefont {Singh}, \citenamefont {Saini},\
  and\ \citenamefont {Tripathi}}]{maurya_performance_2015}%
  \BibitemOpen
  \bibfield  {author} {\bibinfo {author} {\bibfnamefont {J.~B.}\ \bibnamefont
  {Maurya}}, \bibinfo {author} {\bibfnamefont {Y.~K.}\ \bibnamefont
  {Prajapati}}, \bibinfo {author} {\bibfnamefont {V.}~\bibnamefont {Singh}},
  \bibinfo {author} {\bibfnamefont {J.~P.}\ \bibnamefont {Saini}}, \ and\
  \bibinfo {author} {\bibfnamefont {R.}~\bibnamefont {Tripathi}},\ }\href
  {\doibase 10.1007/s11082-015-0233-z} {\bibfield  {journal} {\bibinfo
  {journal} {Optical and Quantum Electronics}\ }\textbf {\bibinfo {volume}
  {47}},\ \bibinfo {pages} {3599} (\bibinfo {year} {2015})}\BibitemShut
  {NoStop}%
\bibitem [{\citenamefont {Zhu}\ \emph {et~al.}(2011)\citenamefont {Zhu},
  \citenamefont {Cheng},\ and\ \citenamefont
  {Schwingenschl{\"o}gl}}]{zhu_giant_2011}%
  \BibitemOpen
  \bibfield  {author} {\bibinfo {author} {\bibfnamefont {Z.~Y.}\ \bibnamefont
  {Zhu}}, \bibinfo {author} {\bibfnamefont {Y.~C.}\ \bibnamefont {Cheng}}, \
  and\ \bibinfo {author} {\bibfnamefont {U.}~\bibnamefont
  {Schwingenschl{\"o}gl}},\ }\href {\doibase 10.1103/PhysRevB.84.153402}
  {\bibfield  {journal} {\bibinfo  {journal} {Physical Review B}\ }\textbf
  {\bibinfo {volume} {84}},\ \bibinfo {pages} {153402} (\bibinfo {year}
  {2011})}\BibitemShut {NoStop}%
\bibitem [{\citenamefont {Chu}\ \emph {et~al.}(2014)\citenamefont {Chu},
  \citenamefont {Schmidt}, \citenamefont {Pu}, \citenamefont {Wang},
  \citenamefont {{\"O}zyilmaz}, \citenamefont {Takenobu},\ and\ \citenamefont
  {Eda}}]{chu_charge_2014}%
  \BibitemOpen
  \bibfield  {author} {\bibinfo {author} {\bibfnamefont {L.}~\bibnamefont
  {Chu}}, \bibinfo {author} {\bibfnamefont {H.}~\bibnamefont {Schmidt}},
  \bibinfo {author} {\bibfnamefont {J.}~\bibnamefont {Pu}}, \bibinfo {author}
  {\bibfnamefont {S.}~\bibnamefont {Wang}}, \bibinfo {author} {\bibfnamefont
  {B.}~\bibnamefont {{\"O}zyilmaz}}, \bibinfo {author} {\bibfnamefont
  {T.}~\bibnamefont {Takenobu}}, \ and\ \bibinfo {author} {\bibfnamefont
  {G.}~\bibnamefont {Eda}},\ }\href {http://dx.doi.org/10.1038/srep07293}
  {\bibfield  {journal} {\bibinfo  {journal} {Scientific Reports}\ }\textbf
  {\bibinfo {volume} {4}},\ \bibinfo {pages} {7293} (\bibinfo {year}
  {2014})}\BibitemShut {NoStop}%
\bibitem [{\citenamefont {Scholz}\ \emph {et~al.}(2013)\citenamefont {Scholz},
  \citenamefont {Stauber},\ and\ \citenamefont
  {Schliemann}}]{scholz_plasmons_2013}%
  \BibitemOpen
  \bibfield  {author} {\bibinfo {author} {\bibfnamefont {A.}~\bibnamefont
  {Scholz}}, \bibinfo {author} {\bibfnamefont {T.}~\bibnamefont {Stauber}}, \
  and\ \bibinfo {author} {\bibfnamefont {J.}~\bibnamefont {Schliemann}},\
  }\href {\doibase 10.1103/PhysRevB.88.035135} {\bibfield  {journal} {\bibinfo
  {journal} {Physical Review B}\ }\textbf {\bibinfo {volume} {88}},\ \bibinfo
  {pages} {035135} (\bibinfo {year} {2013})}\BibitemShut {NoStop}%
\bibitem [{\citenamefont {Kechedzhi}\ and\ \citenamefont
  {Abergel}(2014)}]{kechedzhi_weakly_2014}%
  \BibitemOpen
  \bibfield  {author} {\bibinfo {author} {\bibfnamefont {K.}~\bibnamefont
  {Kechedzhi}}\ and\ \bibinfo {author} {\bibfnamefont {D.~S.~L.}\ \bibnamefont
  {Abergel}},\ }\href {\doibase 10.1103/PhysRevB.89.235420} {\bibfield
  {journal} {\bibinfo  {journal} {Physical Review B}\ }\textbf {\bibinfo
  {volume} {89}},\ \bibinfo {pages} {235420} (\bibinfo {year}
  {2014})}\BibitemShut {NoStop}%
\bibitem [{\citenamefont {Steinhoff}\ \emph {et~al.}(2014)\citenamefont
  {Steinhoff}, \citenamefont {R{\"o}sner}, \citenamefont {Jahnke},
  \citenamefont {Wehling},\ and\ \citenamefont
  {Gies}}]{steinhoff_influence_2014}%
  \BibitemOpen
  \bibfield  {author} {\bibinfo {author} {\bibfnamefont {A.}~\bibnamefont
  {Steinhoff}}, \bibinfo {author} {\bibfnamefont {M.}~\bibnamefont
  {R{\"o}sner}}, \bibinfo {author} {\bibfnamefont {F.}~\bibnamefont {Jahnke}},
  \bibinfo {author} {\bibfnamefont {T.~O.}\ \bibnamefont {Wehling}}, \ and\
  \bibinfo {author} {\bibfnamefont {C.}~\bibnamefont {Gies}},\ }\href {\doibase
  10.1021/nl500595u} {\bibfield  {journal} {\bibinfo  {journal} {Nano Letters}\
  }\textbf {\bibinfo {volume} {14}},\ \bibinfo {pages} {3743} (\bibinfo {year}
  {2014})}\BibitemShut {NoStop}%
\bibitem [{\citenamefont {Liang}\ and\ \citenamefont
  {Yang}(2015)}]{liang_carrier_2015}%
  \BibitemOpen
  \bibfield  {author} {\bibinfo {author} {\bibfnamefont {Y.}~\bibnamefont
  {Liang}}\ and\ \bibinfo {author} {\bibfnamefont {L.}~\bibnamefont {Yang}},\
  }\href {\doibase 10.1103/PhysRevLett.114.063001} {\bibfield  {journal}
  {\bibinfo  {journal} {Physical Review Letters}\ }\textbf {\bibinfo {volume}
  {114}},\ \bibinfo {pages} {063001} (\bibinfo {year} {2015})}\BibitemShut
  {NoStop}%
\bibitem [{\citenamefont {Liu}\ \emph {et~al.}(2013)\citenamefont {Liu},
  \citenamefont {Shan}, \citenamefont {Yao}, \citenamefont {Yao},\ and\
  \citenamefont {Xiao}}]{liu_three-band_2013}%
  \BibitemOpen
  \bibfield  {author} {\bibinfo {author} {\bibfnamefont {G.-B.}\ \bibnamefont
  {Liu}}, \bibinfo {author} {\bibfnamefont {W.-Y.}\ \bibnamefont {Shan}},
  \bibinfo {author} {\bibfnamefont {Y.}~\bibnamefont {Yao}}, \bibinfo {author}
  {\bibfnamefont {W.}~\bibnamefont {Yao}}, \ and\ \bibinfo {author}
  {\bibfnamefont {D.}~\bibnamefont {Xiao}},\ }\href {\doibase
  10.1103/PhysRevB.88.085433} {\bibfield  {journal} {\bibinfo  {journal}
  {Physical Review B}\ }\textbf {\bibinfo {volume} {88}},\ \bibinfo {pages}
  {085433} (\bibinfo {year} {2013})}\BibitemShut {NoStop}%
\bibitem [{\citenamefont {R{\"o}sner}\ \emph {et~al.}(2015)\citenamefont
  {R{\"o}sner}, \citenamefont {{\c S}a{\c s}{\i}o{\u g}lu}, \citenamefont
  {Friedrich}, \citenamefont {Bl{\"u}gel},\ and\ \citenamefont
  {Wehling}}]{rosner_wannier_2015}%
  \BibitemOpen
  \bibfield  {author} {\bibinfo {author} {\bibfnamefont {M.}~\bibnamefont
  {R{\"o}sner}}, \bibinfo {author} {\bibfnamefont {E.}~\bibnamefont {{\c S}a{\c
  s}{\i}o{\u g}lu}}, \bibinfo {author} {\bibfnamefont {C.}~\bibnamefont
  {Friedrich}}, \bibinfo {author} {\bibfnamefont {S.}~\bibnamefont
  {Bl{\"u}gel}}, \ and\ \bibinfo {author} {\bibfnamefont {T.~O.}\ \bibnamefont
  {Wehling}},\ }\href {\doibase 10.1103/PhysRevB.92.085102} {\bibfield
  {journal} {\bibinfo  {journal} {Physical Review B}\ }\textbf {\bibinfo
  {volume} {92}},\ \bibinfo {pages} {085102} (\bibinfo {year}
  {2015})}\BibitemShut {NoStop}%
\bibitem [{\citenamefont {Roth}\ \emph {et~al.}(2014)\citenamefont {Roth},
  \citenamefont {K{\"o}nig}, \citenamefont {Fink}, \citenamefont
  {B{\"u}chner},\ and\ \citenamefont {Knupfer}}]{roth_electron_2014}%
  \BibitemOpen
  \bibfield  {author} {\bibinfo {author} {\bibfnamefont {F.}~\bibnamefont
  {Roth}}, \bibinfo {author} {\bibfnamefont {A.}~\bibnamefont {K{\"o}nig}},
  \bibinfo {author} {\bibfnamefont {J.}~\bibnamefont {Fink}}, \bibinfo {author}
  {\bibfnamefont {B.}~\bibnamefont {B{\"u}chner}}, \ and\ \bibinfo {author}
  {\bibfnamefont {M.}~\bibnamefont {Knupfer}},\ }\href {\doibase
  10.1016/j.elspec.2014.05.007} {\bibfield  {journal} {\bibinfo  {journal}
  {Journal of Electron Spectroscopy and Related Phenomena}\ }\textbf {\bibinfo
  {volume} {195}},\ \bibinfo {pages} {85} (\bibinfo {year} {2014})}\BibitemShut
  {NoStop}%
\bibitem [{\citenamefont {Kresse}\ and\ \citenamefont
  {Hafner}(1993)}]{kresse_textitab_1993}%
  \BibitemOpen
  \bibfield  {author} {\bibinfo {author} {\bibfnamefont {G.}~\bibnamefont
  {Kresse}}\ and\ \bibinfo {author} {\bibfnamefont {J.}~\bibnamefont
  {Hafner}},\ }\href {\doibase 10.1103/PhysRevB.47.558} {\bibfield  {journal}
  {\bibinfo  {journal} {Physical Review B}\ }\textbf {\bibinfo {volume} {47}},\
  \bibinfo {pages} {558} (\bibinfo {year} {1993})}\BibitemShut {NoStop}%
\bibitem [{\citenamefont {Kresse}\ and\ \citenamefont
  {Furthm{\"u}ller}(1996)}]{kresse_efficiency_1996}%
  \BibitemOpen
  \bibfield  {author} {\bibinfo {author} {\bibfnamefont {G.}~\bibnamefont
  {Kresse}}\ and\ \bibinfo {author} {\bibfnamefont {J.}~\bibnamefont
  {Furthm{\"u}ller}},\ }\href {\doibase 10.1016/0927-0256(96)00008-0}
  {\bibfield  {journal} {\bibinfo  {journal} {Computational Materials Science}\
  }\textbf {\bibinfo {volume} {6}},\ \bibinfo {pages} {15} (\bibinfo {year}
  {1996})}\BibitemShut {NoStop}%
\bibitem [{\citenamefont {Friedrich}\ \emph {et~al.}(2010)\citenamefont
  {Friedrich}, \citenamefont {Bl{\"u}gel},\ and\ \citenamefont
  {Schindlmayr}}]{friedrich_efficient_2010}%
  \BibitemOpen
  \bibfield  {author} {\bibinfo {author} {\bibfnamefont {C.}~\bibnamefont
  {Friedrich}}, \bibinfo {author} {\bibfnamefont {S.}~\bibnamefont
  {Bl{\"u}gel}}, \ and\ \bibinfo {author} {\bibfnamefont {A.}~\bibnamefont
  {Schindlmayr}},\ }\href {\doibase 10.1103/PhysRevB.81.125102} {\bibfield
  {journal} {\bibinfo  {journal} {Physical Review B}\ }\textbf {\bibinfo
  {volume} {81}},\ \bibinfo {pages} {125102} (\bibinfo {year}
  {2010})}\BibitemShut {NoStop}%
\bibitem [{_ju(2014)}]{_juelich_2014}%
  \BibitemOpen
  \href {http://www.flapw.de/pm/index.php?n=Main.AboutFleur} {\enquote
  {\bibinfo {title} {The {Juelich} {FLEUR} project},}\ } (\bibinfo {year}
  {2014})\BibitemShut {NoStop}%
\bibitem [{\citenamefont {Mostofi}\ \emph {et~al.}(2008)\citenamefont
  {Mostofi}, \citenamefont {Yates}, \citenamefont {Lee}, \citenamefont {Souza},
  \citenamefont {Vanderbilt},\ and\ \citenamefont
  {Marzari}}]{mostofi_wannier90:_2008}%
  \BibitemOpen
  \bibfield  {author} {\bibinfo {author} {\bibfnamefont {A.~A.}\ \bibnamefont
  {Mostofi}}, \bibinfo {author} {\bibfnamefont {J.~R.}\ \bibnamefont {Yates}},
  \bibinfo {author} {\bibfnamefont {Y.-S.}\ \bibnamefont {Lee}}, \bibinfo
  {author} {\bibfnamefont {I.}~\bibnamefont {Souza}}, \bibinfo {author}
  {\bibfnamefont {D.}~\bibnamefont {Vanderbilt}}, \ and\ \bibinfo {author}
  {\bibfnamefont {N.}~\bibnamefont {Marzari}},\ }\href {\doibase
  10.1016/j.cpc.2007.11.016} {\bibfield  {journal} {\bibinfo  {journal}
  {Computer Physics Communications}\ }\textbf {\bibinfo {volume} {178}},\
  \bibinfo {pages} {685} (\bibinfo {year} {2008})}\BibitemShut {NoStop}%
\end{thebibliography}%

\end{document}